\documentclass{iau} 
\usepackage{graphicx}
\usepackage{natbib}
\usepackage{aas_macros}

\title[Global Evolution of Solar Magnetic Fields and Prediction of Activity Cycles] 
{Global Evolution of Solar Magnetic Fields and Prediction of Activity Cycles}

\author[Irina N. Kitiashvili]   
{Irina N. Kitiashvili}

\affiliation{NASA Ames Research Center, Moffett Field, MS 258-6,
	Mountain View, USA
	\\ email: {\tt irina.n.kitiashvili@nasa.gov}}

\pubyear{2020}
\volume{354}  
\jname{Solar and Stellar Magnetic Fields: Origins and Manifestations}
\editors{A.G. Kosovichev, K. Strassmeier \& M. Jardine, eds.}

\begin{document}
\maketitle

\begin{abstract}
Prediction of solar activity cycles is challenging because physical processes inside the Sun involve a broad range of multiscale dynamics that no model can reproduce and because the available observations are highly limited and cover mostly surface layers. Helioseismology makes it possible to probe solar dynamics in the convective zone, but variations in differential rotation and meridional circulation are currently available for only two solar activity cycles. It has been demonstrated that sunspot observations, which cover over 400 years, can be used to calibrate the Parker-Kleeorin-Ruzmaikin dynamo model, and that the Ensemble Kalman Filter (EnKF) method can be used to link the modeled magnetic fields to sunspot observations and make reliable predictions of a following activity cycle. However, for more accurate predictions, it is necessary to use actual observations of the solar magnetic fields, which are available only for the last four solar cycles. 
In this paper I briefly discuss the influence of the limited number of available observations on the accuracy of EnKF estimates of solar cycle parameters, the criteria to evaluate the predictions, and application of synoptic magnetograms to the prediction of solar activity. 
\end{abstract}

\section{Introduction}
Physics-based solar activity forecasts require knowledge of subsurface and surface evolution of both large-scale flows and magnetic fields. However, at present we have access to measurements of subsurface flow and surface magnetic fields only for the last few solar cycles, which makes calibration of existing dynamo models, as well as the development of new models, very challenging. There have been many attempts to develop models and data analysis techniques to understand the nature of the global dynamics of the Sun and improve the accuracy of activity forecasts \citep[e.g.][]{Dikpati2007,Cameron2007,Choudhuri2007,Kitiashvili2008a,Jouve2011,Karak2013,Upton2018,Jiang2018,Macario-Rojas2018,Covas2019,Labonville2019}. Sunspot number data, available for over 400-years, qualitatively capture variations in the toroidal magnetic field component. This property allows us to use the sunspot number as a proxy of the toroidal field and to calibrate a low-dimensional dynamo model \citep{Kitiashvili2009}. 

Because of inaccuracy in both models and observations, a data assimilation approach has been used to correct model solutions according to corresponding observational data, to estimate uncertainties, to obtain an improved description of the current state of solar activity, and to build a prediction of future states. This analysis is performed using the Ensemble Kalman Filter method \citep{Evensen1997}, which is one of the best methods for applying data assimilation to non-linear problems \citep{Kalnay2002}. It has been applied to assimilate the available synoptic magnetic field observations into the dynamo model and to made estimates for the upcoming Cycle 25 \citep{Kitiashvili2020b} .

\begin{figure}[b]
	\begin{center}
		\includegraphics[width=5.2in]{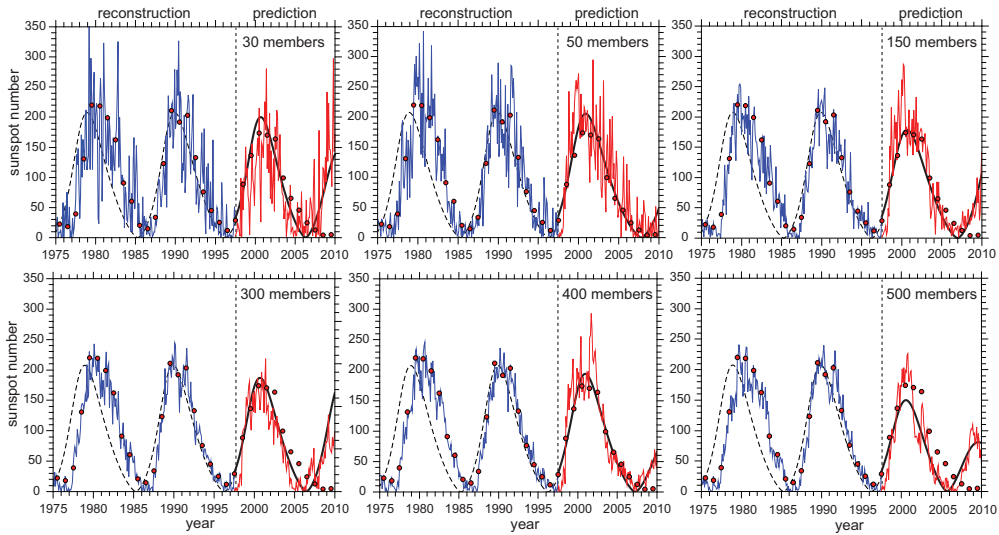} 
		\caption{Test prediction of Solar Cycle 23 for 30, 50 150, 300, 400, and 500 ensemble members. The black curves show the periodic dynamo solution (dashed curves), corrected according to the available observations, to obtain new initial conditions. The exact model solution calculated using the updated initial conditions for the prediction phase is shown by black curves. Blue curves show the ensemble mean model solution corrected according to observations using the EnKF method assuming different sizes of the ensemble. Red curves show the estimated `prediction' of SC23. Black dots show the actual observational data.}
		\label{ens_members}
	\end{center}
\end{figure}

\section{Ensemble Kalman Filter method for solar activity prediction}
The Ensemble Kalman Filter method performs a statistical analysis of possible activity states (so-called ensemble members) and allows us to estimate magnetic field evolution more accurately by taking into account uncertainties in observations as well as potential model errors. To perform the data assimilation analysis, we use a low-mode approximation \citep{Weiss1984} to the mean-field Parker-Kleeorin-Ruzmaikin (PKR) dynamo model in the form of a non-linear dynamical system  \citep{Kitiashvili2009,Kitiashvili2020b}:
\begin{eqnarray}
\frac{{\rm d} A}{{\rm d} t}&=& D B- A, \nonumber \\
\frac{{\rm d} B}{{\rm d} t}&=&{\rm i} A - B,  \\
\frac{{\rm d} \alpha_{m}}{{\rm d} t}&=&-\nu \alpha_{m} - D \left[
B^{2}-\lambda A^2 \right], \nonumber
\end{eqnarray}
where $A$, $B$, $\alpha_{m}$, and $t$ are non-dimensional variables for poloidal field vector-potential, toroidal field strength, magnetic helicity, and time respectively; $D$ is a non-dimensional dynamo number,  $\lambda=$Rm$^{-2}$, where Rm is an effective magnetic Reynolds number,  and $\nu$ is the ratio of characteristic turbulence time-scales \citep{Kleeorin1982}. The reduced dynamo model describes the evolution of the mean global toroidal and poloidal field components and the magnetic helicity, for a set of parameters $D$, $\nu$, and  $\lambda$, and the initial conditions for the initial model state. 
We assume a relationship between the toroidal magnetic field and the sunspot number in the form suggested by \cite{Bracewell1988}, $W = C B^{3/2}$, where $W$ is the sunspot number, and $C$ is a normalization constant.

\begin{figure}[b]
	\begin{center}
		\includegraphics[width=5.2in]{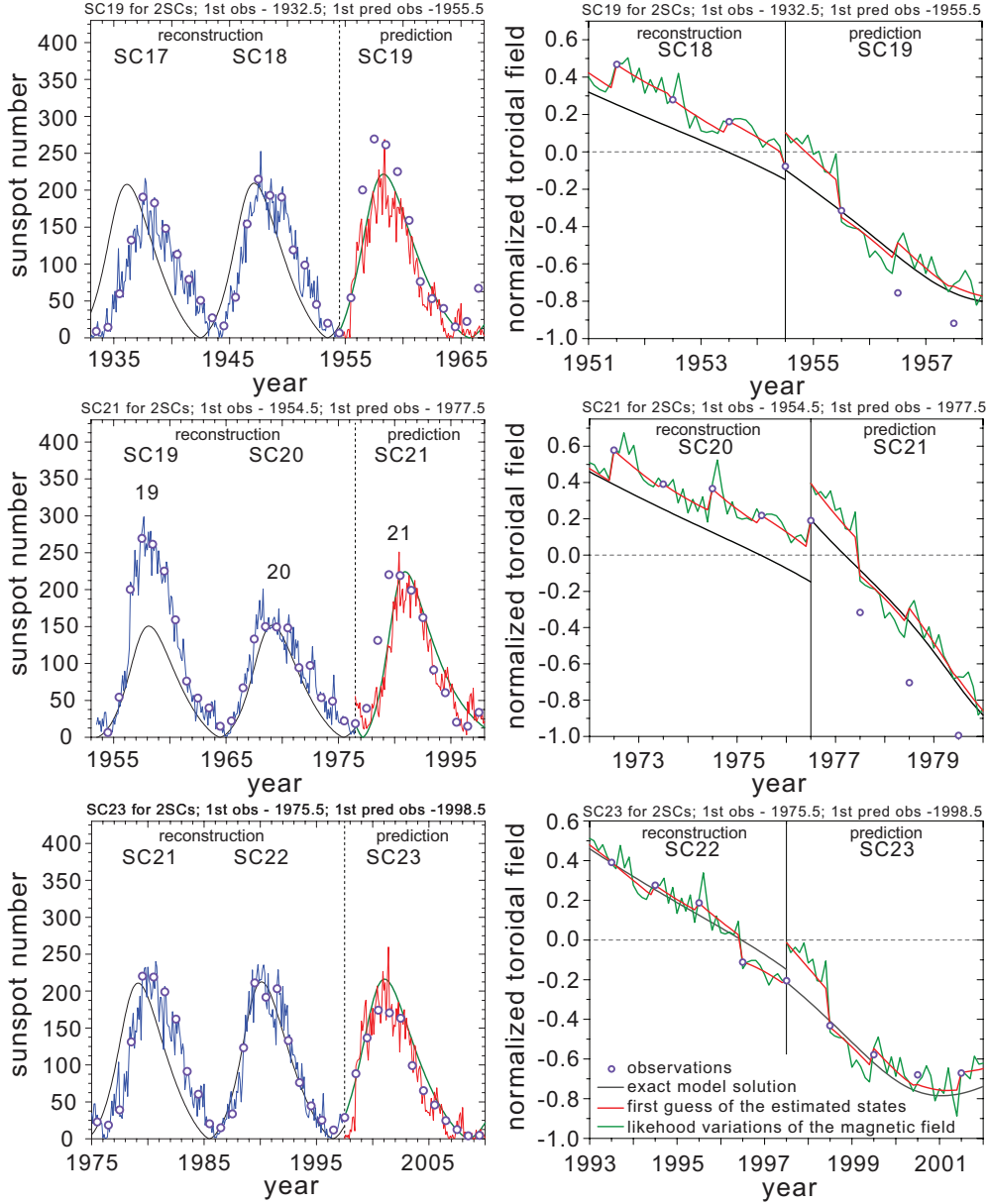} 
		\caption{Predictions of Solar Cycles 19, 21, and 23 using only the annual sunspot number observations for the two preceding solar cycles. Transition between the reconstruction and prediction phases is shown for the toroidal magnetic field component in the right panels.}
		\label{short}
	\end{center}
\end{figure}

The EnKF method has been used for correcting the periodic solution of the model given by (2.1) to predict Solar Cycle 24 (SC24) using the observational data up to the preceding minimum of the solar activity \citep{Kitiashvili2008a}. In addition, \cite{Kitiashvili2016} investigated the potential for early forecast of the next solar cycle using data up to the time of the polar field reversals during the preceding cycle. However, there are several uncertainties in this method. For instance, because of the statistical approach used to estimate the global activity states, the size of the statistical ensemble can affect the resulting solutions. Figure~\ref{ens_members} shows reconstruction of Solar Cycle 23 (SC23) starting from the preceding activity minimum and assuming different sizes of ensemble. In the case of a small ensemble size, the reconstructed sunspot number variations for Cycles 21 and 22 (blue curves) and predictions for Cycle 23 (red curves) are significantly noisier than for larger ensembles. As the size of the ensemble increases, the noise of the solutions decreases. Testing ensemble size effects for several solar cycles allowed us to conclude that, for this problem, using 300 ensemble members is optimal for describing stochastic variations relative to the mean annual sunspot numbers. 

\section{Influence of short series of observational data on prediction accuracy} 
Modern observational data (such as magnetograms) are available only for the last few cycles. Therefore, it is important to investigate the influence of short time-series on prediction accuracy.
Investigation of limited observational time series was performed for solar cycles 19 -- 24 using the annual sunspot number data. Initially, data assimilation was performed using  observations for nine cycles, and then the observational data was sequentially removed cycle-by-cycle and the ability of the procedure to reconstruct a target cycle was evaluated.
Figure~\ref{short} shows test predictions for cycles 19, 21, and 23, when observations of only two preceding cycles were used in the EnKF analysis, performed for 300 ensemble members. The resulting model solutions have been evaluated by using the following criteria: 1) the signs of the last available observation (for toroidal field) and the corresponding model solution should be same; 2) the exact model solution for the prediction phase must be consistent with the model solution for the reconstruction phase (no solution flattening, jumps or 'bumps', but the solution may shift according to new initial condition); 3) the corrected solution (first guess estimate) at the initial moment of time during the prediction phase should not be greater than the best-estimate variations of the toroidal field; and 4) the phase discrepancy between the exact model solution and observations should not be greater than 2 years \citep{Kitiashvili2020a}. 
After the evaluation, we compared the predicted evolution with the actual observations. It was found that the first two criteria are the most important for accuracy estimation. The third criterion has a weaker correlation with the accuracy and is primarily used to make a choice among different solutions when the other criteria are satisfied. We consider this criterion only for the toroidal magnetic field because only sunspot observations are used in the analysis. 

\section{Application of synoptic magnetograms to predict global solar activity}
\subsection{Magnetic field observations}
Our tests with short sunspot number series showed the ability of the EnKF method based on the PKR dynamo model to make, in some cases, a reliable prediction even when observations are available for only two preceding cycles. This makes it reasonable to apply the data assimilation methodology to synoptic magnetograms available from Kitt Peak Observatory \citep{Harvey1980,Worden2000}, the SOLIS instrument \citep{Keller2003}, and SOHO/MDI and SDO/HMI instruments \citep{Scherrer1995,Scherrer2012} for the last four solar cycles, i.e., from 1976 (Carrington rotation 1645) to 2019 (Carrington rotation 2216).

\begin{figure}[b]
	\begin{center}
		\includegraphics[width=5.2in]{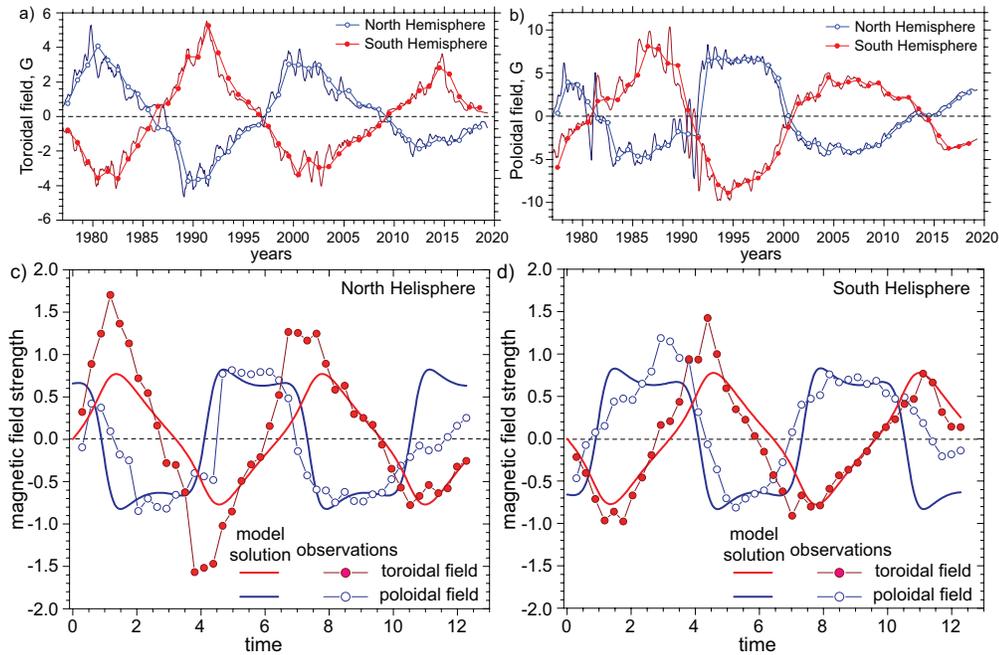} 
		\caption{Temporal variations of the toroidal (panel a) and poloidal fields (panel b) in the northern (blue curves) and southern hemispheres (red curves). Panels c and d: Time-series of the annual toroidal (red dots) and poloidal (blue) field observations calibrated to the corresponding periodic dynamo solutions (thick curves) for the northern (panel a) and southern hemispheres (b). The magnetic fields and time units are non-dimensional.}
		\label{mf}
	\end{center}
\end{figure}

Decomposition of the synoptic magnetograms into toroidal and poloidal field components is challenging due to the difficulty of finding a unique solution, especially from only a line-of-sight magnetic field component. To simplify the magnetic field decomposition problem we assume that the high-latitude magnetic field (above the active latitudes) characterizes the poloidal field component and that the unsigned flux in the active latitudes corresponds to the toroidal field. This assumption is acceptable for the 1D model with some level of uncertainty, because it requires estimates of the relative behavior of the field components. 
To account for toroidal field reversals, the sign of the estimated toroidal field is prescribed according to the Hale polarity law. Figure~\ref{mf}a shows the variation with time in the magnitude of the estimated toroidal field for each hemisphere. The time-series of the estimated toroidal and poloidal fields are averaged over 1-year intervals and are shown by circles for each hemisphere in Figure~\ref{mf}b. Thin curves show the unsmoothed variations of the field components for reference. 

\begin{figure}[t]
	\begin{center}
		\includegraphics[width=5in]{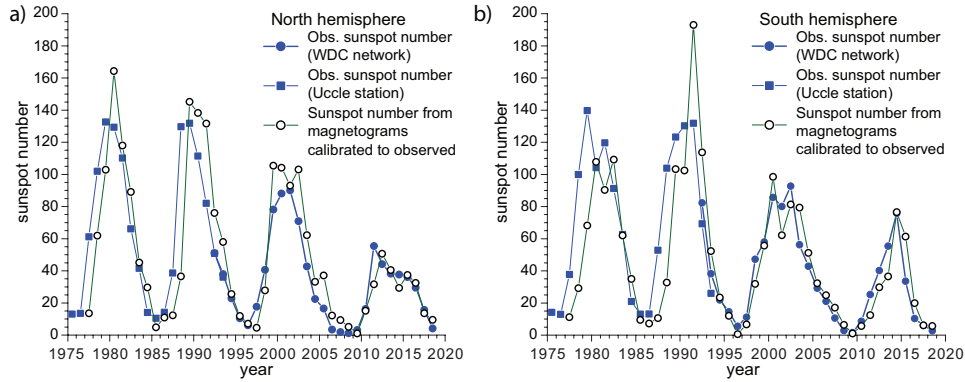} 
		\caption{Comparison of the observed annual sunspot number time-series from the WDC-SILSO network (blue circles, http://www.sidc.be/silso/datafiles), and Uccle station (blue rectangles) with the calibrated annual sunspot number estimated from the synoptic magnetograms (green circles) for: a) northern hemisphere and b) southern hemisphere.}
		\label{MagSN}
	\end{center}
\end{figure}

The resulting annual observations have been normalized to match the model periodic solutions for the toroidal and poloidal fields (Fig.~\ref{mf}~c,~d) for each hemisphere. Normalization for the poloidal field was chosen for best agreement with the field variation amplitude. For the toroidal field, the normalization is performed relative to the last observed solar cycle, following the approach of \cite{Kitiashvili2008a}. Figure~\ref{mf}c shows an example of the toroidal component of the magnetic field calibration in the model solutions for the prediction of SC25. Traditionally, solar activity cycles are characterized by the sunspot number; the toroidal field can be converted to the sunspot number with a corresponding normalization. A comparison of the observed hemispheric sunspot number and that estimated from the synoptic magnetograms is shown in Figure~\ref{MagSN}. It is important to note that discrepancies between the observed sunspot numbers and those estimated from magnetograms can increase the uncertainty in estimates of the sunspot cycle strength. 

\subsection{Solar activity prediction based on the synoptic magnetograms}
In this section we examine the predictive capabilities of the EnKF method as applied to synoptic magnetograms for Solar Cycles 23 and 24 and make a forecast for the upcoming Cycle 25.
The forecasting procedure has been tested for SC23 using synoptic magnetograms obtained during two previous cycles, SC21 and SC22 from 1977.5 to 1996.5. Because the synoptic observations started in the rising phase of SC21, we added two synthetic observations for 1976.5 and 1975.5 corresponding to the solar minimum between SC20 and SC21. Using the EnKF procedure, the periodic model solution is corrected according to the annual observations of the toroidal and poloidal fields for the corresponding hemispheres. The additional model variables, e.g. magnetic helicity, for which observations are not available, are generated from the model solution with an imposed 10\% noise. Comparison of the model prediction with the actual toroidal field variations shows good agreement for both hemispheres up to the SC23 maximum. After the maximum, the predicted toroidal field in the northern hemisphere quickly deviates from the observed evolution. In the southern hemisphere, deviations of the predicted toroidal field become significant two years after the SC23 maximum.  The predicted evolution of the poloidal field in both hemispheres quickly deviates from the actual data after the first prediction year.

\begin{figure}[b]
	\begin{center}
		\includegraphics[width=5in]{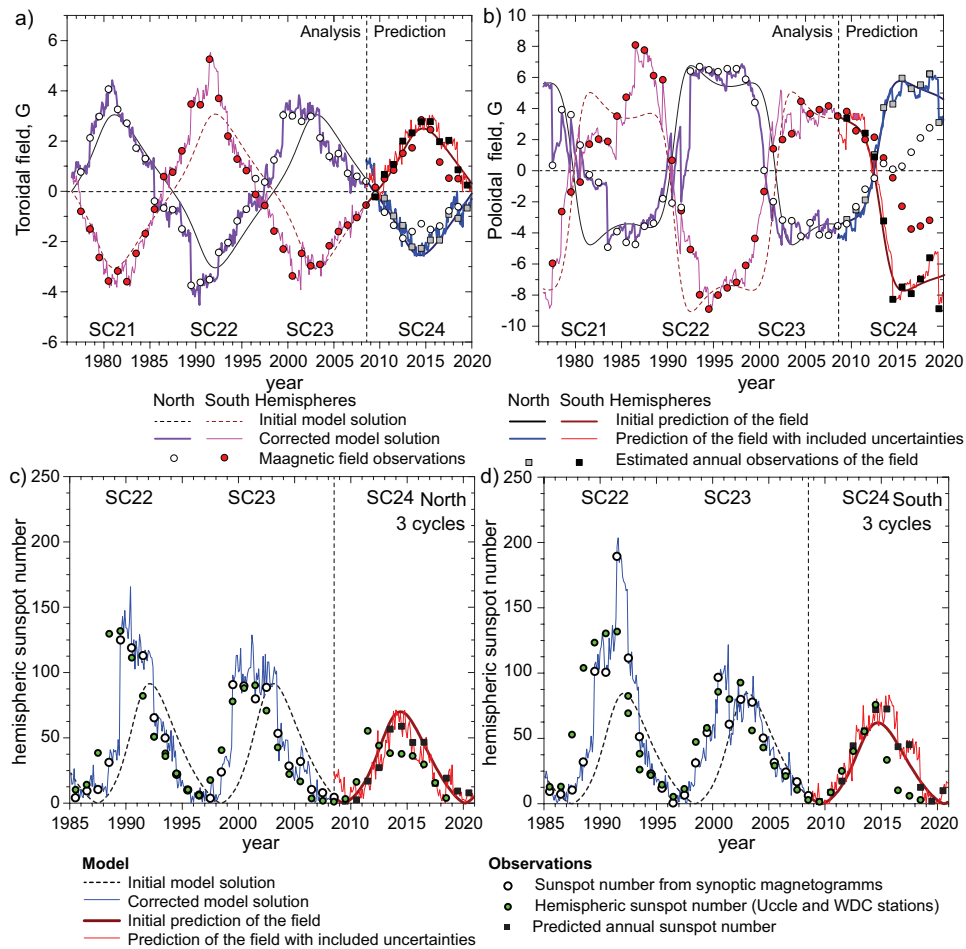} 
		\caption{Predictions for the mean toroidal (panel a) and poloidal (b) fields and sunspot number variations (panels c and d) in the northern and southern hemispheres during Cycle 24. Vertical dashed lines indicate the prediction start time.}
		\label{SC24}
	\end{center}
\end{figure}

\begin{figure}[b]
	\begin{center}
		\includegraphics[width=5in]{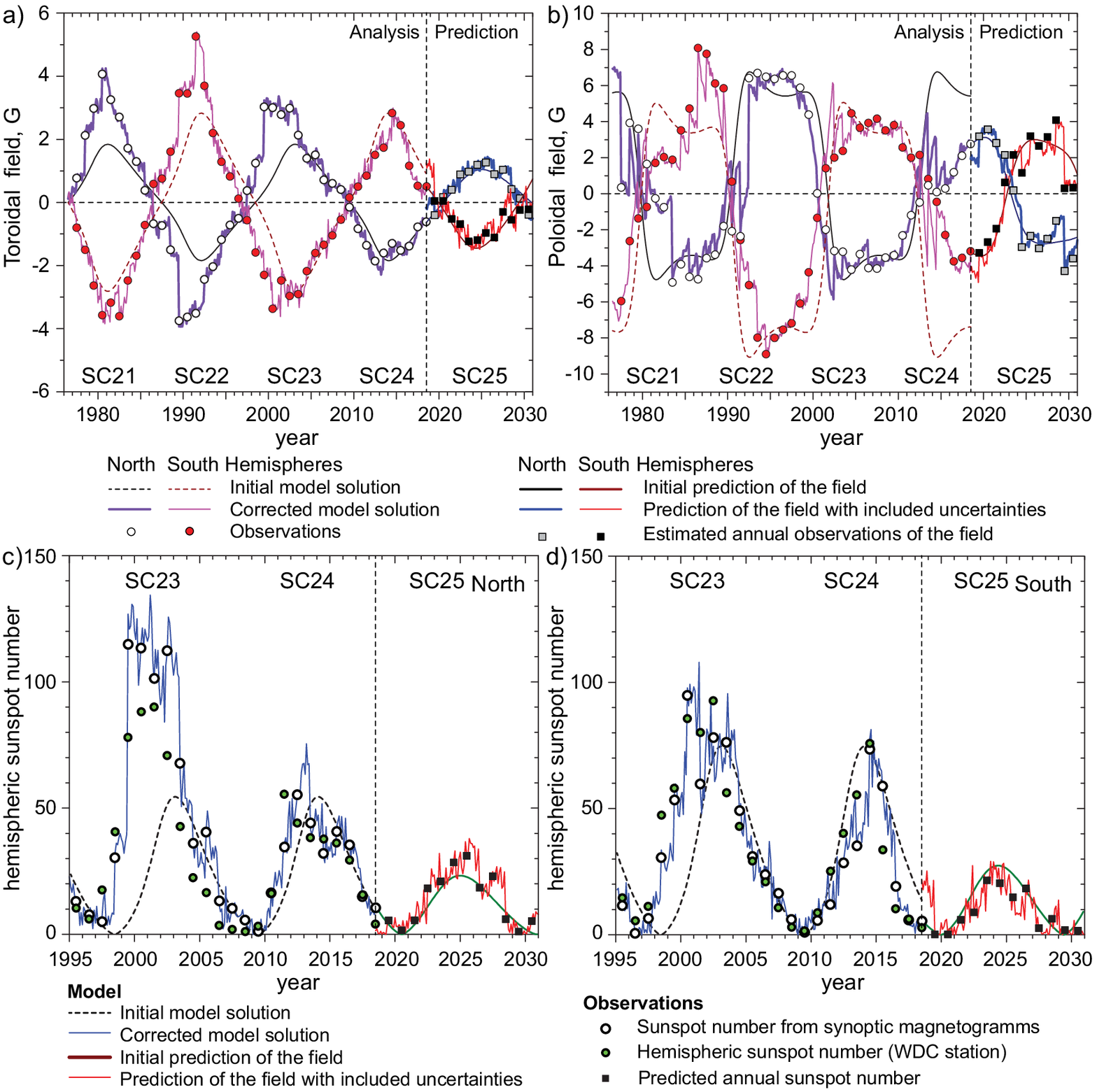} 
		\caption{Predictions for Solar Cycle 25 of the mean toroidal (panel a) and poloidal (b) fields and sunspot number variations (panels c and d) in the northern and southern hemispheres based on field observations for four solar cycles. Vertical dashed lines indicate the prediction start time.}
		\label{SC25}
	\end{center}
\end{figure}

The test prediction for Solar Cycle 24 was performed by using assimilation of  observational data for the previous two and three solar cycles \citep{Kitiashvili2020b}. In the three cycle case, we used the available magnetic field measurements from 1977.5 to 2008.5 to predict SC24  (Fig.~\ref{SC24}). The predicted evolution of the toroidal field is in good agreement for both hemispheres, although there are some discrepancies during the solar maximum in the northern hemisphere. 
Increasing discrepancies between the predicted and observed toroidal fields during the decay phase of solar activity in the southern hemisphere are expected because of the step-like variations of the predicted toroidal field evolution (red thin curve, Fig.~\ref{SC24}a) and the sunspot number (red thin curve, Fig.~\ref{SC24}d). This behavior indicates accumulation of errors during the analysis and, in general, gives us a warning that the forecast quality is potentially low. This effect previously was discussed by \cite{Kitiashvili2008a} for assimilation of the sunspot-number time-series. 
Accuracy of the poloidal field prediction (Fig.~\ref{SC24}b) is good for up to 3 years and provides a correct prediction for the time of the polar field reversals. After this, the predicted and observed field components quickly diverge. The sunspot number estimates (Fig.~\ref{SC24}) show good agreement with the actual data for both hemispheres. Some deviations in the shape of the predicted activity cycles are expected, and this reflects restrictions of the dynamo model formulation. The total sunspot number maximum is slightly overestimated, but in general the prediction results show a good agreement for the whole solar cycle.

To perform a prediction of upcoming Solar Cycle 25, all four solar cycles of synoptic magnetic field data from 1977 to 2019 have been used. Figure~\ref{SC25} shows the predictions for the toroidal (panel a) and poloidal (panel b) fields and the sunspot number (panel c and d) in both hemispheres. As expected, the forecast for the toroidal fields (and the sunspot number) is more accurate for the northern hemisphere than for the southern hemisphere because of smaller discrepancies between the model solution and observations at the end of Cycle 24.
The model solutions show strong variation in the toroidal fields near and after 2026.5 (red curves, Figs~\ref{SC25}a, c, d). These strong variations indicate that prediction uncertainties significantly increase after 2026.5. Thus, the sunspot number prediction for SC25 in the northern hemisphere is about 30 (that is $\sim 50$\% weaker than SC24) with an error of 15 -- 20\% and about 25 for southern hemisphere ($\sim 65$\% weaker than SC24) with error 25 -- 30\%. The solar maximum is expected during 2024 -- 2026 in the northern hemisphere, and during 2024 -- 2025 in the southern hemisphere.

\section{Discussion and conclusions}
With very limited information about the evolution and structure of the magnetic fields and flows in the interior, reconstruction of the current state of global solar activity and prediction of future activity is a challenging task. The limited observational data restrict our ability to build accurate global models. The long series of sunspot number observations only provide a rough estimate of the global toroidal magnetic field, and, though the synoptic magnetograms carry more information about the toroidal and poloidal field evolution, they are available for only four cycles.

Nevertheless, correlations between the surface magnetic fields and the sunspot number variations allow us to test ideas of how to improve long-term solar activity predictions by developing new models and data analysis techniques and invoking data assimilation and machine learning approaches. Using the sunspot time series we were able to calibrate the low-order mean-field PKR dynamo model \citep{Kitiashvili2009} using an approximate relationship between sunspot number and the global toroidal magnetic field strength, identify criteria to evaluate the prediction quality \citep{Kitiashvili2020a}, and demonstrate the potential of the EnKF data assimilation method to make a reliable prediction of future solar activity using the data for only three preceding sunspot cycles. In some cases, the derived criteria gave us a warning that the prediction results may be not accurate. More work needs to be done for developing quantitative criteria for the forecasting accuracy.

Application of the Ensamble Kalman Filter method to predict solar cycle variations using synoptic magnetograms shows certain limitations because the magnetograms are available for only four activity cycles. In addition, there is no accurate procedure to uniquely decompose the line-of-sight magnetic field measurements onto poloidal and toroidal field components. Also, there is no one-to-one correspondence between the observed sunspot numbers (Fig.~\ref{MagSN}) and those estimated from magnetograms; this adds some bias in the interpretation of the prediction results. Nevertheless, the data assimilation approach combined with synoptic magnetogram data allowed us to make predictions for the next solar cycle. According to this analysis, Solar Cycle 25 will be weaker than the current cycle and will start after an extended solar minimum during 2019 -- 2021. The maximum of activity will occur in 2024 -- 2025 with a sunspot number at the maximum of about $50\pm 15$ (for the v2.0 sunspot number series) with an error estimate of ~30\%. The Solar Cycle will start in the southern hemisphere in 2020 and reach maximum in 2024 with a sunspot number of $\sim 28$ ($\pm 10$\%). Solar activity in the northern hemisphere will be delayed for about 1 year (with an error of $\pm 0.5$~year) and reach maximum in 2025 with a sunspot number of $\sim 23 \pm 5$ ($\pm 21$\%).

The presented results encourage future development of the data assimilation methodology for more detailed dynamo models and more complete data sets, and, in particular, development of cross-analysis of different data sources to characterize the global dynamics of the Sun.

{\bf Acknowledgment.} The work is supported by NSF grant AGS-1622341.


\bibliographystyle{aa}

\end{document}